\documentclass[preprint,showpacs,superscriptaddress,aps]{revtex4-1}
\usepackage{amssymb}
\usepackage{amsmath,amssymb,graphicx}
\usepackage{graphicx}% Include figure files
\usepackage{dcolumn}% Align table columns on decimal point
\usepackage{bm}% bold math
\def\be{\begin{equation}}
\def\ee{\end{equation}}
\def\bea{\begin{eqnarray}}
\def\eea{\end{eqnarray}}

\def\lsim{\raise0.3ex\hbox{$\;<$\kern-0.75em\raise-1.1ex\hbox{$\sim\;$}}}
\def\gsim{\raise0.3ex\hbox{$\;>$\kern-0.75em\raise-1.1ex\hbox{$\sim\;$}}}

\newcommand{\nn}{\nonumber}

\usepackage{axodraw4j}
\usepackage{pstricks}
\begin{document}

\bigskip

\vspace{2cm}
\title{Lepton number violation in top quark and neutral B meson decays}
\vskip 6ex
\author{David Delepine}
\email{delepine@fisica.ugto.mx}
\affiliation{Division de Ciencias e Ingenier\'ias,  Universidad de Guanajuato, Campus Leon,
 C.P. 37150, Le\'on, Guanajuato, M\'exico.}
\author{G. L\'opez Castro}
\email{glopez@fis.cinvestav.mx}
\affiliation{Departamento de F\'\i sica, Centro de Investigaci\'on y de Estudios Avanzados, Apartado Postal 14-740, 07000 M\'exico D.F., M\'exico}
\author{N. Quintero}
\email{nquintero@fis.cinvestav.mx}
\affiliation{Departamento de F\'\i sica, Centro de Investigaci\'on y de Estudios Avanzados, Apartado Postal 14-740, 07000 M\'exico D.F., M\'exico}

\bigskip

\bigskip

\begin{center}
\begin{abstract}
 Lepton number violation (LNV) can be induced by Majorana neutrinos in four-body decays of the neutral $B$ meson and the top quark. We study the effects of Majorana neutrinos in these $|\Delta L|=2$ decays in an scenario where a single heavy neutrino can enhance the amplitude via the resonant mechanism. Using current bounds on heavy neutrino mixings, the most optimistic branching ratios turn out to be at the level of $10^{-6}$ for $\bar{B}^0\to D^+e^-e^-\pi^+$ and $t\to bl^+l^+W^-$  decays. Searches for these LNV decays at future facilities can provide complementary constraints on masses and mixings of Majorana neutrinos.
\end{abstract}
\end{center}

\pacs{11.30.Fs, 13.20.He, 14.60.Pq, 14.60.St, 14.65.Ha}
\maketitle
\bigskip
%\baselinestretch{1.5}

\section{ Introduction}

\bigskip

After being established that neutrinos are massive and mixed particles \cite{Fukuda:1998mi,Wendell:2010md,Ambrosio:2003yz}, one of the most interesting current issues in flavor physics is to elucidate if neutrinos are Dirac or Majorana fermions \cite{Mohapatra:2005wg}. If neutrinos turn out to be Majorana particles, important consequences as lepton number-violating (LNV) processes\cite{Pontecorvo:1957qd,Pontecorvo:1967fh} and further sources of CP violation become possible \cite{Bilenky:1980cx,Schechter:1980gr,Langacker:1986jv,Doi:1980yb}. Searches for LNV processes (where the lepton number is violated in two units, $\Delta L=2$) in dedicated low energy experiments as neutrinoless double beta decays, have led to very strong constraints on the effective mass of light Majorana neutrinos \cite{KlapdorKleingrothaus:2000sn,Aalseth:2002rf,Arnaboldi:2008ds,Daraktchieva:2009mn,Rodejohann:2011mu} since the rates for these processes are driven by the effective mass parameter $\langle m_{ee} \rangle$ \cite{effmass}. On the other hand, very restrictive bounds on $\langle m_{ll'} \rangle$ can be obtained by combining neutrino oscillation data \cite{GonzalezGarcia:2010er} , cosmological bounds \cite{Lesgourgues:2006nd,Sekiguchi:2009zs,Acero:2008rh,GonzalezGarcia:2010un} and tritium beta decay \cite{Otten:2008zz}. Interestingly, these sub-eV bounds on the scale of effective Majorana masses are at the sensitivity reaches of current experimental projects \cite{Osipowicz:2001sq,Ardito:2005ar,Aalseth:2005mn,Ikeda:2005yh}.

  As it has been extensively discussed by many authors, the existence of very light neutrinos may find a natural explanation by means of heavy neutrinos via the see-saw mechanism \cite{Georgi:1974sy,Minkowski:1977sc,gellman, yanagida, glashow,Weinberg:1979sa,Mohapatra:1979ia}. Heavy neutrinos naturally appears in some extensions of the Standard Model and may play an important role in cosmology and various particle physics and astrophysical processes \cite{Atre:2005eb,Atre:2009rg}. The possibility to observe the effects of heavier neutrinos, accessible in the kinematical range of current experiments, is very exciting as they can induce large rates for $\Delta L=2$ decays through the mechanism of resonant enhancement \cite{Atre:2009rg}. Indeed, the appearance of sterile neutrinos with masses in the range of hundreds of MeV's to a few GeV's is possible in scenarios of dynamical electroweak symmetry breaking as shown for instance in \cite{intermediate,Appelquist:2003hn,Appelquist:2003hd}. By means of the resonant mechanism, neutrinos with these intermediate mass scales can produce an enhancement in the three-body $\Delta L=2$ decays of pseudoscalar mesons $M_1^+ \to l^+l^+M_2^-$ and the tau lepton $\tau^- \to l^+M_1^-M_2^-$; these decay processes have been extensively studied by many authors \cite{Atre:2009rg,Littenberg:1991ek,three-body,Zhang:2010um,Helo:2010cw,Cvetic:2010rw} in
the cases where final state hadrons can be pseudoscalar or vector mesons. So far, some experimental uppers bounds have been reported in refs. \cite{pdg,He:2005iz,Rubin:2010cq,babar:2011hb}; very recently, by using 36 pb${-1}$ of integrated luminosity, the LHCb collaboration has been able to derive upper limits for LNV charged $B$ meson decays $B(B^+ \to K^-(\pi^-)\mu^+\mu^+)<5.8(5.4)\times 10^{-8}$ \cite{lhcb}. These studies are expected to be extended by the LHCB experiment by including the $B^+ \to D^-_{(s)}\mu^+\mu^+,\  \bar{D}^0\mu^+\mu^+\pi^-$ decay modes \cite{lhcb}, which toghether with similar analyses that can be performed at the SuperB Flavor Factories \cite{superB} makes very attractive the studies of LNV  $B$ meson decays. Similarly, like-sign dileptons may be produced via the resonance enhancement mechanism in four-body decays of top quarks and W gauge bosons, as it has been investigated for instance in Refs. \cite{top1,Si:2008jd,Kovalenko:2009td}.

In the present paper we consider the four-body decays of neutral $B$ mesons, $\bar{B}^0 \to D^+l^-l^-\pi^+$ with $l=e,\ \mu$, in the favored scenario of resonant neutrino enhancement. The dynamics of this four-body decay involves the transition $B\to D$ form factors and is different from the one driving the three-body decays of mesons and tau leptons which involve the meson decay constants. To our knowledge, these $\Delta L=2$ decays of neutral $B$ mesons have not been investigated before neither from a theoretical nor from an experimental point of view. In addition, we also consider and update the analogous four-body $t \to bl^+l^+W^-$ decays ($l=e,\ \mu,\ \tau$), which was previously studied in \cite{top1}, since one naively expects it can be largely enhanced due to the resonances in the virtual $W$ boson and heavy neutrino exchanges.

\bigskip

\section{ Four-body $\Delta L=2$ decays of heavy flavors}

  The Feynman diagrams that describe the LNV decays of the top quark and neutral $B$ meson are shown in Figures 1 and 2, respectively. Contributions containing the properly antisymmetrized contributions due to the exchange of identical leptons in the final state must be added to these diagrams.

Following previous studies \cite{top1,Atre:2009rg},  we consider a model with three left-handed SU(2) lepton doublets $L_{aL}^T= (\nu_a,\ l_a)_L$, ($a=1,2,3$), and $n$ right-handed singlets $N_{bR}$ $(b=1,2, \cdots n)$. In the basis of mass eigenstates, the charged current interactions of leptons are given by \cite{Atre:2009rg}:
\be
{\cal L}_{l}^{\rm ch}= - \frac{g}{\sqrt{2}} W^+_{\mu}
\left(\sum_{l=e}^{\tau}\sum_{m=1}^{3}V_{lm}\bar{\nu}_m\gamma^{\mu}P_Ll+  \sum_{l=e}^{\tau}\sum_{m=1}^{n}U_{lm}\overline{N^c_m}\gamma^{\mu}P_L l\right) + {\rm h. \ c.}
\ee
where $P_L=(1-\gamma_5)/2$ is the left-handed chirality operator,  $g$ is the $SU(2)_L$ gauge coupling, $\psi^c\equiv C\bar{\psi}^T$ is the charge conjugated spinor, and $V_{lm}$ ($U_{lm}$) denotes the light (heavy) neutrino mixings; the subscript $m$ refers to the mass eigenstate basis entering the diagonalized Majorana mass term for neutrinos \cite{Atre:2009rg}:
\be
{\cal L}_m^{\nu}=-\frac{1}{2} \left(\sum_{m=1}^{3}m^{\nu}_m \overline{\nu_{mL}}\nu^c_{mR}+ \sum_{m=4}^{n}m^N_{m}\overline{N^c_{mL}} N_{mR} \right)+ {\rm h.\ c.}\ .
\ee
In the phenomenological applications of the present paper, we will assume that only one heavy neutrino with mass $m_N$ and charged current couplings $U_{lN}$ to leptons, dominates the decay amplitudes via the resonant enhancement mechanism.

\begin{figure}
  % Requires \usepackage{graphicx}
  \includegraphics[width=5cm]{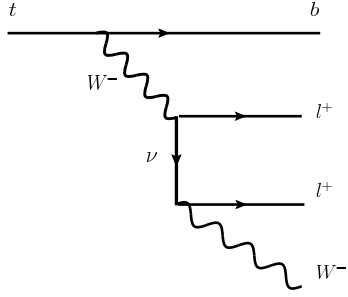}
%\vspace{-18.0cm}
  \caption{Feynman graph for the $t\to bl^+l^+W^-$ decay.}\label{fig1}
\end{figure}

\begin{figure}
  % Requires \usepackage{graphicx}
  \includegraphics[width=8cm]{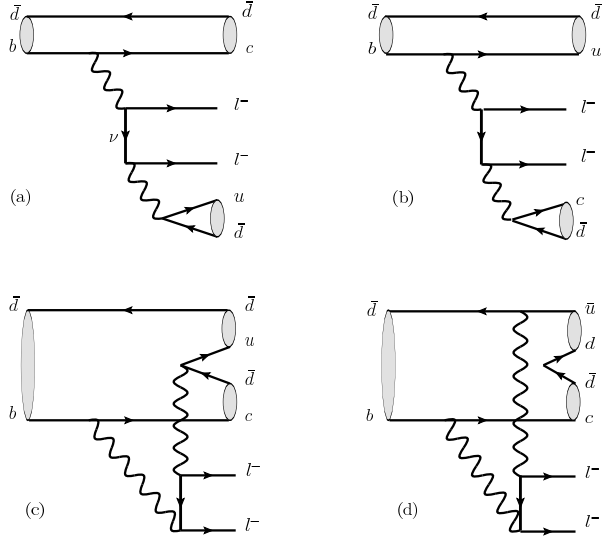}
%\vspace{-18.0cm}
  \caption{Feynman graph for the $\Delta L=2$ neutral $\bar{B}$ meson decay.}\label{fig2}
\end{figure}

The kinematics of four-body decays can be described in terms of five independent variables. In our convention of momenta and masses they are defined as  
\be
P(p, M) \to P_1(p_1, m_1)P_2(p_2,m_2)P_3(p_3,m_3)P_4(p_4,m_4) \nonumber 
\ee
 with $p^2=M^2$ and $p_i^2=m_i^2$. We choose the set of independent variables as $\{s_{12}, s_{34}, \theta_1, \theta_3, \phi\}$ which have the following geometrical meaning \cite{Pais:1968zz} (see Figure 3):
   \begin{itemize}
    \item $s_{12}\equiv (p_1+p_2)^2$, is the invariant-mass of particles 1 and 2; 
    \item $s_{34} \equiv (p_3+p_4)^2$, is the invariant-mass of particles 3 and 4;
    \item $\theta_1$ ($\theta_3$), is the angle between the three-momentum of particle 1 (particle 3) with respect to the direction of $\vec{p}_{12}\equiv \vec{p}_1+\vec{p}_2$ (respectively, $\vec{p}_{34}\equiv \vec{p}_3+\vec{p}_4$) defined in the rest frame of the decaying particle;
    \item $\phi$ is the angle between the planes defined by particles $(1,2)$ and $(3,4)$ also in the rest frame of the decaying particle.
  \end{itemize}
 
\begin{figure}
  % Requires \usepackage{graphicx}
  \includegraphics[width=9cm]{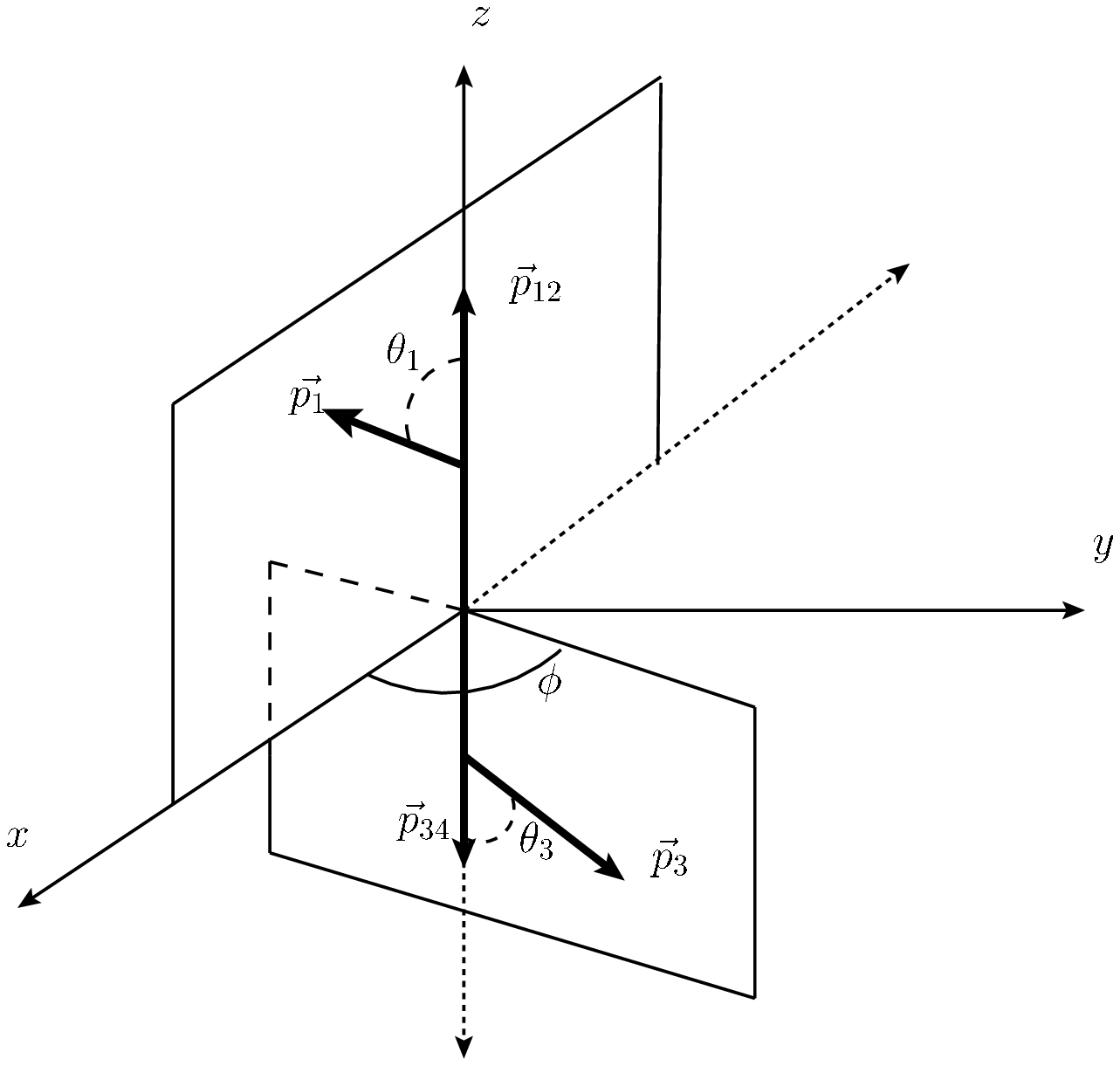}
%\vspace{-18.0cm}
  \caption{Kinematics of four-body decays in the rest frame of the decaying particle, $\sum_{i=1}^{4}\vec{p_i}=0$. We have defined $\vec{p}_{ij}=\vec{p}_i+\vec{p}_j$, such that $\vec{p}_{12}+\vec{p}_{34}=0.$}\label{fig3}
\end{figure}

With this choice of kinematics, the differential decay rate in the rest frame of the decaying particle can be written as:
\be
d\Gamma=\frac{X\beta_{12}\beta_{34}}{4(4\pi)^6M^3}\overline{|{\cal M}|^2}\cdot \frac{1}{n!}ds_{12}ds_{34}d\cos \theta_1 d\cos \theta_3 d\phi \ ,
\ee
where $\beta_{12}$ ($\beta_{34}$) is the velocity of particle 1 (particle 3) in the center of mass frame of particles 1 and 2 (3 and 4) and $X=\lambda^{1/2}(M^2, s_{12}, s_{34})/2$, with $\lambda(x,y,z)\equiv x^2+y^2+z^2-2xy-2xz-2yz$. Finally, $\overline{|{\cal M}|^2}$ is the spin-averaged squared amplitude of the four-body decay. In our case of two identical leptons in the final state, $n=2$.

\bigskip

\section{ $B \to D ll\pi$ decays}

Let us first consider the $\bar{B}^0(p)\to D^+(p_1)l^-(p_2)l^-(p_3)\pi^+(p_4)$ decays (see Figure 2), where $p_i$ denote the four-momenta of final state particles. In the range of neutrino masses $m_N$ where the resonance effects dominate the decay amplitude, the diagrams of Figure 2(c) and 2(d) will give very small contributions. In addition, we note that diagram 2(b) is suppressed with respect to 2(a) due to smaller Cabibbo-Kobayashi-Maskawa (CKM) factors ($|V_{ub}V_{cd}/(V_{cb}V_{ud})|\sim 0.02$). Therefore, we keep the diagram shown in 2(a) as the dominant contribution.

The properly antisymmetrized decay amplitude is given by:
\be
{\cal M}
= G_F^2 V_{cb}V_{ud}\langle D(p_1)|\bar{c}\gamma^{\mu}b|B(p)\rangle \cdot\bar{u}(p_2)\left[{\cal P}_N(p_2)\gamma_{\mu}\gamma_{\nu}+ {\cal P}_N(p_3)\gamma_{\nu}\gamma_{\mu}\right]u^c(p_3)\left(if_{\pi}p_4^{\nu}\right)
\ee
where $V_{ij}$ denotes the $ij$ entry of the CKM quark mixing matrix, $G_F$ is the Fermi constant and $f_{\pi}=130.4$ MeV is the $\pi^+$ decay constant.

In the above expression we have defined the factor
\be
{\cal P}_N(p_i)=\frac{U_{lN}^2m_{N}}{(Q-p_i)^2-m_{N}^2+im_{N}\Gamma_{N}}\ ,
\ee
where  $Q=p-p_1=p_2+p_3+p_4$ is the momentum transfer and $U_{lN}$ denotes the heavy neutrino mixing defined in Section 2. $\Gamma_N$ represents the decay width of the heavy neutrino which depends on the decay channels that can be opened at the mass $m_N$; it allows to keep finite the amplitude when $(Q-p_i)^2=m_{N}^2$.  As it was pointed out above, in this paper we will assume that only one heavy neutrino $N$ falls in the resonance region of the $B$ meson and top quark decays, thus it will give the dominant contribution to the decay amplitude. The mixings of the heavy neutrino with the three charged leptons will be taken as the currently most restrictive bounds as reported in Ref. \cite{delAguila:2008pw}
\be
{\rm Set I:}\ \ \ |U_{eN}|^{2} < 3 \times 10^{-3}, \ \ |U_{\mu N}|^{2} < 3 \times 10^{-3}, \ \ |U_{\tau N}|^{2} < 6 \times 10^{-3}\ .
\ee

The hadronic matrix element in Eq. (4) is given by:
\be
\langle D^+(p_1)|\bar{c}\gamma_{\mu}b| \bar{B}^0(p)\rangle =\left( (p+p_1)_{\mu}- \frac{m_B^2-m_D^2}{t}Q_{\mu}\right) F_1(t)+ \frac{m_B^2-m_D^2}{t}Q_{\mu}F_0(t)\ ,
\ee
where $t=Q^2$. For the purposes of a numerical evaluation, we will use two common parametrizations of the form factors $F_{1,0}(t)$, namely the one provided by the Wirbel-Stech-Bauer (WSB) model \cite{WSB}:
\be
F_1^{\rm WSB}(t)=\frac{F_1^{\rm WSB}(0)}{1-t/m_{1^-}^2},\ \ \ \ F_0^{\rm WSB}(t)=\frac{F_0^{\rm WSB}(0)}{1-t/m_{0^+}^2}\ ,
\ee
where $F_1^{\rm WSB}(0)=F_0^{\rm WSB}(0)=0.69$, $m_{1^-}= 6.34$ GeV and $m_{0^+}=6.8$ GeV \cite{WSB} and, just for comparison, we will use also the parametrization provided by the covariant light front (CLF) model \cite{CLF}:
\be
F_1^{\rm CLF}(t)=\frac{F_1^{\rm CLF}(0)}{1-a_1(t/m_B^2)+b_1(t/m_B^2)^2},\ \ \ \
F_0^{\rm CLF}(t)=\frac{F_0^{\rm CLF}(0)}{1-a_0(t/m_B^2)+b_0(t/m_B^2)^2}\ ,
\ee
where $F_1^{\rm CLF}(0)=F_0^{\rm CLF}(0)=0.67$, $a_1=1.25$, $b_1=0.39$, $a_0=0.65$ and $b_0=0.0$ \cite{CLF}.

\begin{figure}
  % Requires \usepackage{graphicx}
  \includegraphics[width=13cm]{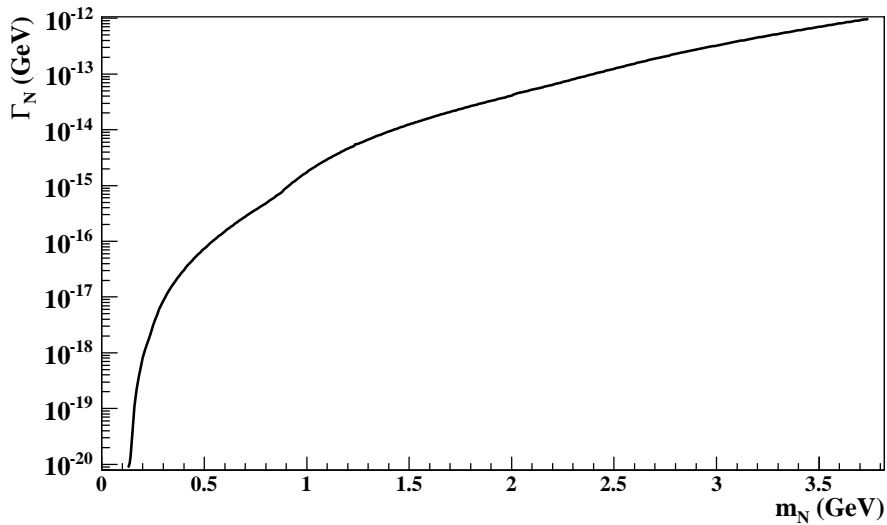}
%\vspace{-18.0cm}
  \caption{Neutrino decay width for neutrino masses relevant to produce resonant enhancement in $\bar{B}^0 \to D^+ l^-l^-\pi^+$ decays.}\label{fig4}
\end{figure}

  The decay width of the intermediate neutrino state is obtained by adding up the contributions of all the neutrino decay channels that can be opened at the mass $m_N$ \cite{Atre:2009rg}:
\be
\Gamma_N =\sum_f \Gamma(N\to f)\theta (m_N-\sum_{i} m_{f_i})\ ,
\ee
where $m_{f_i}$ in the argument of the step function are the masses of the final state particles in the neutrino decay channel $f$. The dominant decay modes of the neutrino in the range of masses relevant for resonant $B$ meson decays are the following: $l^{\mp}P^{\pm}$, $\nu_l P^0$, $l^{\mp}V^{\pm}$, $\nu_l V^0$, $ l_1^{\mp} l_2^{\pm}\nu_{l_2}$, $\nu_{l_1}l_2^-l_2^+$, and $\nu_{l_1}\nu \bar{\nu}$, where $l,\ l_1,\ l_2=e,\ \mu,\ \tau$, and $P$ ($V$) denotes a pseudoscalar (vector) meson state. The expressions for the partial decay rates of these channels can be found in Appendix C of Ref. \cite{Atre:2009rg}.
  We have re-evaluated the decay width of the neutrino which is plotted in Figure 4 for neutrino masses $m_N$ in the range where it can produce a resonant enhancement of the $B$ meson decay amplitude. The decay width $\Gamma_N$ is so tiny that the Narrow Width Approximation (NWA) of Eq. (5),
\be
\lim_{\Gamma_N\to 0}{\cal P}_N(p_i)=-i\pi m_N U_{lN}^2\delta \left((Q-p_i)^2-m_N^2\right)
\ee
 is required to perform the numerical integrations to compute the decay rates from Eq. (3).

In Figure 5 we plot the branching ratios of $\bar{B}^0 \to D^+l^-l^-\pi^+$ decays for the electron (dashed-line) and muon (solid-line) channels as a function of the neutrino mass $m_N$.
%%%%%%
\begin{figure}
  % Requires \usepackage{graphicx}
  \includegraphics[width=13cm]{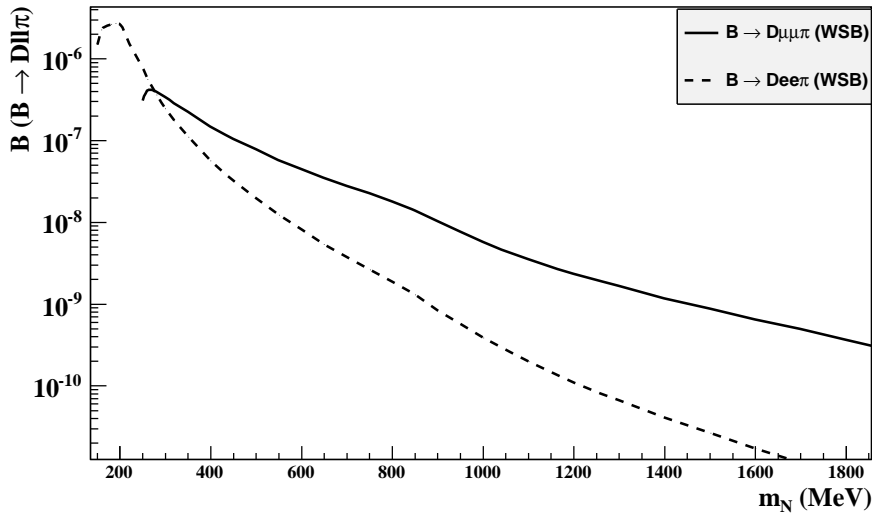}
%\vspace{-18.0cm}
  \caption{Branching ratio of $\bar{B}^0 \to D^+l^-l^-\pi^+$ decay as a function of $m_N$. The dashed (solid) line  corresponds to the electronic (muonic) channel. }\label{fig5}
\end{figure}
%%%%%%%%%%%%
These plots were obtained by using the WSB model \cite{WSB} for the form factors $F_{1,0}(t)$. The branching fractions reach their maximum values for neutrino masses that are close to the threshold for the $m_l+m_{\pi}$ production and they decrease for increasing values of $m_N$.
\begin{table}[t]
\centering
\small
\renewcommand{\arraystretch}{1.1}
\renewcommand{\arrayrulewidth}{0.8pt}
\begin{tabular}{cccccc}
\hline\hline
 &  $\bar{B}^{0} \to D^{+} e^{-}e^{-} \pi^{+}$ &  &  & $\bar{B}^{0} \to D^{+} \mu^{-}\mu^{-} \pi^{+}$ &\\
\hline
$m_{N}$ (MeV) & WSB & CLF & $m_{N}$ (MeV) & WSB & CLF \\
 \hline
170	& 2.6$\times 10^{-6}$ & 3.4$\times 10^{-6}$ & 250 & 3.0$\times 10^{-7}$ & 3.9$\times 10^{-7}$\\
190	& 2.8$\times 10^{-6}$ & 3.6$\times 10^{-6}$ & 270 &4.1$\times 10^{-7}$ & 5.4$\times 10^{-7}$\\
200	& 2.6$\times 10^{-6}$ & 3.4$\times 10^{-6}$ & 300 & 3.4$\times 10^{-7}$ & 4.3$\times 10^{-7}$ \\
220	& 1.5$\times 10^{-6}$ & 2.0$\times 10^{-6}$ & 400 & 1.4$\times 10^{-7}$ & 1.9$\times 10^{-7}$ \\
250	& 7.3$\times 10^{-7}$ & 9.7$\times 10^{-7}$ & 500 & 7.0$\times 10^{-8}$ & 1.0$\times 10^{-7}$ \\
300	& 2.5$\times 10^{-7}$ & 3.3$\times 10^{-7}$ & 600 & 4.0$\times 10^{-8}$ & 6.0$\times 10^{-8}$ \\
\hline\hline
\end{tabular}
\caption{{\small Branching ratios for $\bar{B}^{0} \to D^{+} \ell^{-}\ell^{-} \pi^{+}$ decays using the Set I of the heavy neutrino mixings.  WSB \cite{WSB} and CLF \cite{CLF} refer to the form factor models for the $B \to D$ transition}.}
\end{table}

In Table I we show the largest possible values of the branching ratios of $\bar{B}^0 \to D^+l^-l^-\pi^+$ decays ($l=e,\ \mu$), which correspond to the lower range of  neutrino masses. We have evaluated these results for the two different form factor models mentioned in Eqs. (8) and (9). Although the predictions for the form factors in the two models exhibit large differences in the full range of the momentum transfer $t$, the integrated rates differ only at the level of 30\% for almost all values of neutrino masses, both in the electronic and muonic decay channels. The largest possible values of the branching fractions shown in Table I are of the same order as the ones corresponding branching ratios reported for the $B^{+} \to P^{-}l^+l^+$ decays \cite{pdg} and their study can be useful to get further constraints on the heavy neutrino mixings.

\bigskip

\section{ LNV top quark decay}

  LNV transitions with $\Delta L=2$ has been studied also at higher energies. The $t \to bl_i^+l_j^+W^-$ decay (and its crossed $W^+\to l_i^+l_j^++$ 2 jets channel) has been considered previously in Ref. \cite{top1}; similar top quark decays that also include the final $W$ gauge boson decay into two jets were studied  in \cite{Si:2008jd}. The top decay can be resonantly enhanced if the heavy neutrino mass lies in the range $m_W+m_l \leq m_N \leq m_t-m_b-m_l$. In addition (see Figure 1) we can expect an enhancement of the top quark decay amplitude due to the virtual $W$ boson emitted  from the top quark vertex which can be produced also in a resonant way. As it was emphasized in ref. \cite{top1}, the final state $W$ boson in this $\Delta L=2$ top quark decay has the `wrong' charge signature when compared to the dominant $t \to bW^+$ decay. In this Section we provide an update for this same-sign dilepton decay channel in top quark decay by using the neutrino mixings provided in Eq. (6); furthermore, our results for this decay channel provide a test for the particular kinematics that we use in our calculations and that was described in  Section 2.

For the purposes of comparison with previous results on $\Delta L=2$ top quark decays \cite{top1}, we will evaluate our results using, in addition to the Set I of values given in Eq. (6), the following set of neutrino couplings \cite{Bergmann:1998rg}
\be
{\rm Set II:} \ \ \ |U_{eN}|^{2} < 12 \times 10^{-3},\ \ |U_{\mu N}|^{2} < 9.6 \times 10^{-3}, \ \ |U_{\tau N}|^{2} < 16 \times 10^{-3}\ .
\ee

The decay amplitude for $t(p) \to b(p_1)l^+(p_2)l^+(p_3)W^-(p_4)$ corresponding to the diagram of Figure 1 is given by:
\bea
{\cal M}&=& i\frac{G_F m_W^2}{\sqrt{2}}\left( \frac{g}{\sqrt{2}}\right)V_{tb} \bar{u}(p_1)\gamma_{\rho}(1-\gamma_5)u(p)\cdot D_W^{\rho\mu}(Q)\nn \\
&& \ \ \ \ \times \bar{u}(p_2) \left[ {\cal P}_N(p_2)\gamma_{\mu}\gamma_{\nu}+ {\cal P}_N(p_3)\gamma_{\nu}\gamma_{\mu}\right] u^c(p_3)\cdot \epsilon^{\nu}(p_4)\ ,
\eea
where $D_W^{\rho\mu}(Q)=i(-g^{\rho\mu}+Q^{\rho}Q^{\mu}/m_W^2)/(Q^2-m_W^2+im_W\Gamma_{W})$, with $Q=p-p_1$, denotes the resonant $W$ boson propagator  in the unitary gauge, $\epsilon^{\nu}(p_4)$ is the polarization four-vector of the $W^-$ boson and ${\cal P}_N(p_i)$ was defined in Eq. (5).

  The total decay width of the neutrino for the range of neutrino masses giving rise to the resonance enhancement, is determined from the following set of two-body final states: $N \to l^{\pm}W^{\mp},\ \nu_l Z^0$ and $\nu_l H$. The expressions for the total width by neglecting the charged lepton masses is given by \cite{top1,Atre:2009rg}:
\be
\Gamma_N=\frac{G_F\sum_l |U_{lN}|^2}{8\sqrt{2}\pi m_N^3}\left[2(m_N^2+2m_W^2)X_W+(m_N^2+2m_Z^2)X_Z+m_N^2X_H \right]\ ,
\ee
where $X_i=(m_N^2-m_i^2)^2\theta(m_N-m_i)$ for $i=W^{\pm},\ Z$ and $H$ bosons.
 As long as the neutrino mass increases, the total decay width $\Gamma_N$ also grows because of the neutrino mass dependence and also because new decay channels are opened. For neutrino masses relevant for top quark decays, $\Gamma_N$ is several orders of magnitude larger than for $B$ meson decays and a straightforward evaluation of the five-dimensional integration of Eq. (3) can be done without numerical complications; at the same time, the neutrino width is small enough that it allows also the use of the NWA approximation, Eq. (11), to integrate the phase space.
\begin{figure}
  % Requires \usepackage{graphicx}
  \includegraphics[width=13cm]{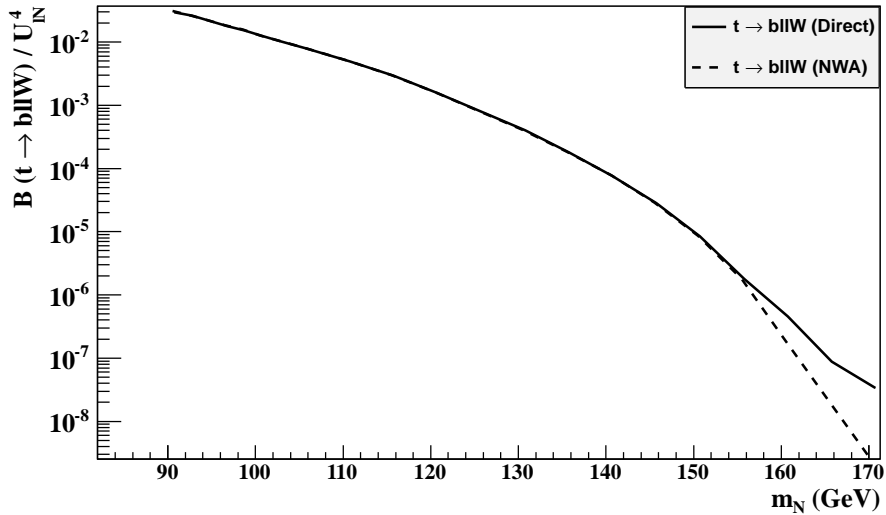}
%\vspace{-18.0cm}
  \caption{Normalized branching ratio of $t \to bl^+l^+W^-$ decay as a function of $m_N$. The solid-line is obtained by a straightforward integration of Eq. (3), and the dashed-line corresponds to the NWA for the neutrino width.}\label{fig6}
\end{figure}

In Figure 6 we plot the branching fraction (normalized to $|U_{lN}|^4$) of $t \to b l^+l^+W^-$ decays as a function of $m_N$. The phase space and the squared amplitudes are almost insensitive to the masses of different leptonic channels in the final state. We use as inputs: $m_t=172.0$ GeV, $m_H=120$ GeV, the leptons and gauge bosons masses given in ref. \cite{pdg}, and the neutrino mixings given in Eq. (6). The solid-line represents the branching ratio that is obtained from the five-dimensional integration of Eq. (3). The dashed-line is the result obtained by using the NWA method. Clearly, the results obtained by using these two methods are almost identical except for very small differences appearing at the upper values of neutrino masses that can produce the resonant enhancement. As it was pointed out above, the results shown in Figure 6 were obtained by using the Set I of neutrino mixings given in Eq. (6). Results for a {\rm new} set of neutrino mixings can be obtained by multiplying the results shown in Figure 6 by $\sum_l |U_{lN}^{\rm Set\ I}|^2/(\sum_l|U_{lN}^{\rm new}|^2)$ .

  \begin{table}[!h]
\centering
\small
\renewcommand{\arraystretch}{1.2}
\renewcommand{\arrayrulewidth}{0.8pt}
\begin{tabular}{cccc}
\hline\hline
&  & Set I &  \\
\cline{2-4}
  $m_{N}$ (GeV) & $ee$ & $\mu\mu$ & $\tau\tau$ \\
 \hline
90 & 0.29 & 0.29 & 1.12 \\
100 & 0.12 & 0.12 &	0.47 \\
110 & 0.05 & 0.05 &	0.19 \\
\hline
 &  & Set II &  \\
\cline{2-4}
  $m_{N}$ (GeV) & $ee$ & $\mu\mu$ & $\tau\tau$ \\
  \hline
90 & 1.48 (1.4) & 0.95 (1.1) & 2.55 (1.9) \\
100 & 0.62 (0.6) & 0.40 (0.5) & 1.08 (0.8) \\
\hline\hline
\end{tabular}
\caption{{\small Branching ratios (in $10^{-6}$ units) for $t \to b \ell^{+} \ell^{+} W^{-}$ decays. Results of Ref. \cite{top1} corresponding to the Set II of neutrino mixings are shown within parenthesis.}}
\end{table}

In Table II we show the branching ratios of $t \to b l^+l^+W^-$ ($l=e,\ \mu, \ \tau)$ decays for a few values of neutrino masses such that the rates have their largest values. For this specific range of neutrino masses, the branching ratios turn out to be of order $10^{-6}\sim 10^{-7}$. For a given Set of neutrino mixings, the results for different leptonic channels differ basically by the rescaling of their fourth power of $U_{lN}$, as it should be. Just for comparison, we have computed the branching ratios by using the Set II of neutrino mixing parameters; our results are compared with those in Ref. \cite{top1} (shown within parenthesis in Table II). Our results and those of reference \cite{top1} have  similar values for the electronic and muonic channels, but they differ in the $\tau\tau$ channel by about 30\%.
 Finally, let us comment that we have evaluated the branching ratios of $t \to bl^+l^+W^-$ for a wider range of the heavy neutrino mass. The normalized branching ratio $B(t\to bllW)/|U_{lN}|^4$ plotted in Figure 6 drops from $1.7\times 10^{-8}$ to $8.3\times 10^{-11}$ when the neutrino mass spans from 200 GeV to 2 TeV. 

We close this section with an estimate of the expected sensitivities to
the signals of our $\Delta L=2$ decays. Using as a (conservative) reference the 
$\sim 450 \times 10^6$ $B\bar{B}$ pairs accumulated by the BABAR collaboration at the 
$\Upsilon(4S)$, we can provide an estimate of the sensitivity to the branching ratio of $\bar{B}^0 \to D^+ l^-l^-\pi^+$ decay at the $B$ factories by using the  $K^-\pi^+\pi^+$ mode to reconstruct the charged $D$ meson. By assuming a 70\% efficiency for the identification
and reconstruction of each of the six charged tracks in the final state one can reach a sensitivity of $\sim 2.0 \times 10^{-7}$ which can test some range of our upper limits for the di-leptonic channels. Of course, this optimistic estimate does not include the combinatorial background for this detection channel, although it can motivate a more detailed study of backgrounds and efficiencies (note also that, under similar assumptions, slightly better sensitivities can be reached using Belle data which is about the double of $B$ meson pairs produced by BABAR). Improved sensitivities can be obtained at the Super Flavor Factories \cite{superB} which are expected to accumulate a larger data set by a factor of 50 to 75 with respect to $B$ factories. Also, as it was mentioned in the Introduction, there are good perspectives at the LHCb Experiment which can provide sensitive improvements on these LNV decays in the dimuon channel based on the analysis already done in the case of LNV searches in $B^+$ meson decays \cite{lhcb}. 
 Regarding the top quark $\Delta L=2$ decays, the sensitivies that can be reached at the Large Hadron Collider (LHC) are not sufficient to test our predictions, as it has been discussed in Ref. \cite{top1}. Only in the eventual case of an upgraded Super-LHC Experiment, which should increase the LHC luminosity by a factor of 10 \cite{Gianotti:2005ak}, one can expect that branching ratios of $10^{-6}$ for $t \to b l^+l^+W^-$ decays could be accessible.

\bigskip

\section{ Conclusions}

  The existence of lepton number-violating transitions with $\Delta L=2$, is considered to be the cleanest manifestation of neutrinos as Majorana particles. Direct searches of these decays in neutrinoless double beta nuclear decays indicate that masses for very light Majorana neutrinos  lie at the sub-eV scale. Similarly, in the framework where only three light neutrinos exist, the Majorana masses of all neutrinos are strongly constrained from current oscillation, cosmological bounds and tritium beta decay.

  In the present paper we have studied the potential of heavy flavor four-body $\Delta L=2$ decays to shed some light on the masses and couplings of heavier Majorana neutrinos. If the masses of such neutrinos produce a resonance enhancement of these heavy flavor decays, the corresponding branching ratios turns out to be, in the most optimistic cases, at the level of $10^{-6}$ for neutral $B$ meson  and top quarks decays if the most restrictive current bounds on neutrino mixings (Set I) are used. These branching fractions of four-body neutral $B$ meson decays are at the level of the upper bounds obtained in experimental studies of $B^{\pm}$ three-body decays. Thus, their searches at current and future experimental facilities can be helpful to provide complementary constrains to the ones derived from three-body decays of $\tau$ leptons and charged $B$ and $D$ mesons.

\bigskip

\subsection*{Acknowledgements}

The authors are grateful to Conacyt (M\'exico) and DAIP project (Guanajuato University) for financial support. They also very grateful to Jose Ben\'\i tez and Carlos Ch\'avez for useful discussions.


\begin{thebibliography}{99}
\bibitem{effmass}
We use the notation $\langle m_{ll'}\rangle \equiv \sum_i V_{li}V_{l'i}m_i$, $V_{li}$ being the corresponding mixings of the three light mass ($m_i$) Majorana neutrinos.
\end{thebibliography}
\end{document}